\documentclass[amsmath,amssymb,preprint]{revtex4}
\usepackage{latexsym}
\usepackage{graphicx}
\usepackage{dcolumn}
\usepackage{bm}
\def\eqnn#1{(\ref{eq:#1})}
\def\plb{{  Phys. Lett. }}
\def\prl{{  Phys. Rev. Lett. }}

\def\figno#1{Fig.~\ref{fig:#1}}

\def\vev#1{\langle#1\rangle}
\begin{document}
\title{Some thermal transport properties of the FPU model with
  quadratic pinning}
\author{
  Kenichiro Aoki\footnote{E--mail:~{\tt ken@phys-h.keio.ac.jp}.
    }}
\affiliation{Hiyoshi Dept. of Physics, Keio University, {\it
    4---1---1} Hiyoshi, Kouhoku--ku, Yokohama 223--8521, Japan}
\begin{abstract}
    Thermal transport properties of the FPU $\beta$ model with a
    quadratic pinning term are investigated for various couplings and
    temperatures. In particular, the size dependence of the thermal
    conductivity, $\kappa\propto L^\alpha$, is studied. $\alpha$ 
    agrees with that of the FPU $\beta$ model (with no pinning) at high
    temperatures but decreases at low temperatures.  This crossover
    behavior occurs at a temperature depending on the strength of the
    quadratic pinning.
\end{abstract}

\vspace{3mm}

\maketitle

Physics of non-equilibrium is of importance to a broad range of issues
in science.
Consequently, non-equilibrium physics has been studied for a long
time, yet a number of basic problems remain. Thermal transport in one
dimensional systems, while seemingly simple, has been an area of
intriguing active research for a long time\cite{fpuRev}.
The FPU model is one of the most classic models in studies of
non-equilibrium physics and a natural model defined on a lattice, so
that it has been studied from various points of view\cite{fpuRev}.  By
virtue of being well studied, many of its properties in the
non-equilibrium steady state are known and has been quite useful as a
guide in studying non-equilibrium physics in various models at finite
temperatures. One of the key intriguing properties of the model is the
lack of bulk behavior in thermal transport. Specifically, the thermal
conductivity of the model is known to behave as $\kappa\propto
L^\alpha$, where $L$ is the size of the system and $\alpha$ ranges
from $1/3$ to
$2/5$\cite{fpuRev,Lepri97,ak-fpu,Narayan02,Lepri03,Pereverzev,MDN07}.

In this work, we study thermal properties of a model which generalizes
the so called FPU $\beta$ model by including a quadratic pinning term
and study its properties, from first principles. The Hamiltonian of
the theory is
\begin{equation}
    \label{eq:hamGeneral}
 H'= \sum_{k=1}^L \left[ {p_k^{\prime 2}\over2m}
  + m\omega^2{(q'_{k+1}- q'_k)^2\over2} + {\mu'q_k^{\prime2}\over2}
  + \beta{(q'_{k+1}- q'_k)^4\over4}\right]
\end{equation}
We study the statistical mechanical properties related to thermal
transport and analyze how the behavior changes with the the physical
parameters in the theory.
FPU model combined with the quartic and quadratic pinning potentials
and with randomness have been studied
recently\cite{WangLi06,Dhar07,LiLi07}. 

Let us explain why we find this model interesting: The FPU model does
not have bulk behavior, as mentioned above. This can be attributed to
the translational symmetry of the theory\cite{Prosen00,fpuRev} and is
seen in a range of similar models. Any pinning term, such as the one
we introduce in \eqnn{hamGeneral} destroys this symmetry, so that the
model may achieve bulk behavior. However, if we ignore the quartic FPU
interaction, the model described by \eqnn{hamGeneral} is none other
than a harmonic model whose exact solutions are known and do not
display sensible bulk behavior\cite{Rieder67,Nakazawa}. So by
combining these models, we have a situation wherein neither model has
bulk behavior, yet the combined model might, which we investigate.
Also, there are arguments as to why the size dependence of the thermal
conductivity is universal in theories with translational
symmetry\cite{Lepri97,Narayan02,Lepri03}. As we remove this symmetry,
we expect that the size dependence will not be the same so that we
have a model in which the size dependence can change naturally with
the parameters of the theory.
An objective here is to obtain insight into the thermal properties of
the model by studying how $\alpha$ changes with the physical
parameters. Furthermore, the model is a fairly simple one generalizing
the well studied FPU model.  As such, knowing the properties of the
model globally in parameter space, we hope, will enhance our
understanding of the thermal transport properties of similar theories.

Let us first reduce the model to its simplest form without any loss in
generality.  Using the rescalings in $q_k,p_k$ and time, we obtain the
following form of the Hamiltonian, similarly to the FPU
model\cite{ak-fpu}.
\begin{equation}
    \label{eq:ham}
    H= \sum_{k=1}^L \left[ {p_k^{2}\over2}
  + {(\nabla q)_k^2\over2} + {\mu q_k^{2}\over2}
  + {(\nabla q)_k^4\over4}\right]
\end{equation}
Here, the variables are related as $q'=\omega \sqrt {m/\beta}\, q,\
p'=\omega^2\sqrt{m^3/\beta}\,p$, $\mu'=m\omega^2\mu$,
$H'=(m^2\omega^4/\beta)H$.  We also adopted the notation $\nabla
q_k\equiv q_{k+1}-q_k$ for convenience. The temperature in the two
formulations are related as $T=\beta T'/(m^2\omega^4)$, and $T$ is
effectively the strength of the FPU coupling, similarly to the FPU
model and the $\phi^4$ theory\cite{ak-fpu,ak-phi}. The model has a
coupling constant, $\mu$, the strength of the quadratic pinning.  As
we can see from the reparametrizations, the quadratic pinning becomes
more important for low temperatures. The size $L$ of the system is a
parameter of the model which enters intrinsically into transport
properties, if the model does not have bulk behavior.  We choose to
fix the quartic coupling and vary $T$, since it seems more natural to
fix the system and vary the temperature, but one may equivalently fix
$T$ and vary the coupling, due to the above rescaling degrees of
freedom.

We now investigate the thermal transport properties of this system by
studying its non-equilibrium steady states. The energy flow in the
system is defined locally as $J_k=-p_k\nabla q_k(1+\nabla q_k^2)$.
The system at
finite temperature has three parameters $\mu,L,T$ and we try to
elucidate the physics behind the dependence on these parameters,
similarly to the case of $\mu=0$ in \cite{ak-fpu}.
Let us briefly describe how the thermal conductivity, $\kappa$, is
computed: Non-equilibrium steady states are numerically constructed
for a given set of $\mu,L$ and $T$ using thermostats generalizing the
Nos\'e--Hoover thermostats,  at the
boundaries\cite{nose-hoover,thermostats}. In this work, we
thermostatted two sites at each end of the system at temperatures
$(T_1^0,T_2^0)$.
Away from these thermostatted boundary points, the behavior of the
system is governed dynamically by the Hamiltonian \eqnn{ham},
including the boundary temperature jumps.
One crucial point is that the thermostats we employ are known to be
able thermalize the harmonic oscillator chain\cite{thermostats}, which
the classic Nos\'e-Hoover thermostats can not\cite{nose-hoover,hoover-harm}. 
 This
is relevant here since as we vary the parameters, the system in some
cases will approach the harmonic model.
The classical equations of motion were integrated numerically using
the Runge--Kutta method with time steps of $0.002$ to $0.02$. The
physical results were checked to be stable against variations in the
size of the time step.  $10^7$ to few times $10^9$ data were averaged
to obtain the physics results. The system sizes used varied from
$L=36$ to $L=4000$.
Fixed boundary conditions are used throughout.  The standard ideal
gas thermometer, $T=\vev{p_k^2}$, was used to measure the temperature
locally.

\begin{figure}[htbp]
    \centering
    \includegraphics[width=8cm,clip=true]{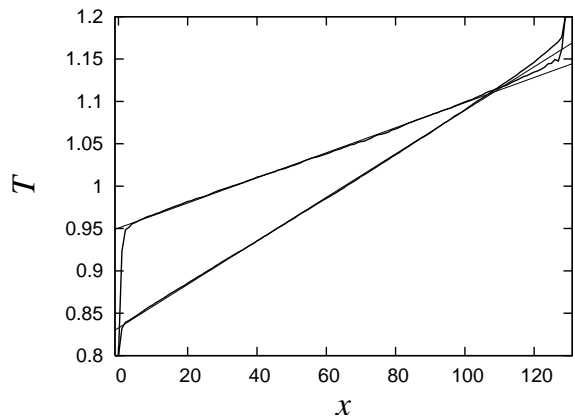}
    \caption{Temperature profiles with $\mu=1$ for
      $(T_1^0,T_2^0)=(0.8,1.2), (0.08,0.12)$. Temperature has been
      rescaled as $2T/(T_1^0+T_2^0)$.  Jumps are larger for the latter
      boundary conditions.  Linear fits to the profiles away from the
      boundaries are also shown but they fit the profile so well that
      they can be seen only close to the boundaries. }
    \label{fig:tProf}
\end{figure}

Thermal conductivity is obtained from Fourier's Law
\begin{equation}
    \label{eq:fourier}
    J=-\kappa\nabla T     
\end{equation}
where $J$ is the heat current.  To obtain $\kappa$, multiple
non-equilibrium states are constructed around the same average
temperature $T$ to confirm that Fourier's law is valid and also to
reduce error. This also allows us to check that the we are not too far
away from equilibrium and that the linear response still holds.  This
is performed for fixed values of $\mu$ and $L$.  Examples of thermal
profiles in non-equilibrium steady states are shown in \figno{tProf}.

In order to obtain $\kappa$ reliably, the following two points need to
be taken care of.  Traditionally, one often obtains $\alpha$ through
the behavior of the heat current $J$ for fixed boundary conditions as
$JL\sim L^{\alpha_J}$.  However, in general, temperature jumps arise
at the boundaries, as seen in \figno{tProf}, and this 
needs to be taken into account\cite{Hatano}.
Due to the jumps, the gradient is not $(T_2^0-T_1^0)/L$, but smaller.
Since the jumps are proportional to the boundary temperature
gradients\cite{pk,ak-jumps}, they will be smaller for larger systems
when the boundary conditions are fixed. Consequently, there is a
tendency for $\alpha_J\geq\alpha$ to hold. A way to avoid this is to
measure the thermal gradient inside the system, away from the
boundaries, as we shall do so here.  This is particularly important
when the system has parameters such that its is close to the harmonic
limit and has large boundary jumps.  The occurrence of large jumps is
consistent with the exact analytic results known for the harmonic
chains, whose thermal profiles consist essentially only of boundary
jumps\cite{Rieder67,Nakazawa}.  Some profiles with their linear fits
for deriving the gradient are shown in \figno{tProf}, which illustrate
this behavior.
Another issue that needs to be addressed is the validity of the linear
response theory. If we stray too far from equilibrium, $J$ will
deviate appreciably from linear response theory. Here, we check that
the linear relation holds and that we keep the gradients to be not too
large for each parameter set.  A priori, there is no rule as to
whether $J$ has to be larger or smaller than the linear response
prediction, but in the FPU model and the $\phi^4$ theory, $|J|$ is
smaller than what is expected from $\kappa\nabla T$\cite{ak-nonEq}.
If the boundary temperature differences are fixed while varying $L$,
$\nabla T$ is larger for smaller systems so that they are further from
equilibrium.  This effect can lead to an underestimation of $\kappa$
for smaller systems, which in turn can also lead an overestimation of
$\alpha$. The problem can be avoided by checking Fourier's law
\eqnn{fourier} or using the knowledge of how large the deviation from
the linear response theory is.
While it will not play an important role here, curvature in the
profiles can also affect the extraction of $\kappa$ in general.

\begin{figure}[htbp]
    \centering
    \includegraphics[width=8cm,clip=true]{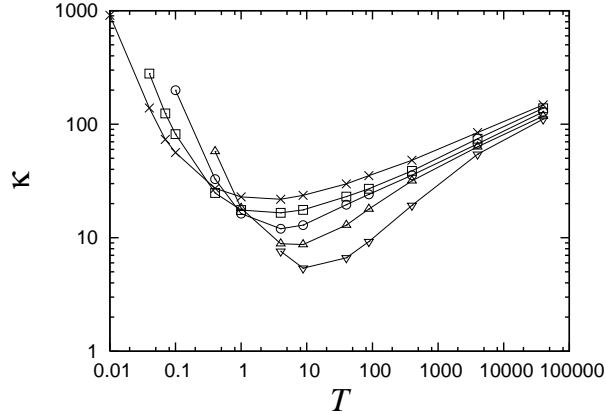}
    \caption{Temperature dependence of $\kappa$ with $L=132$, for
      $\mu=0\ (\times)$, $\mu=1\ (\Box)$, $\mu=4\ (\bigcirc)$,
      $\mu=10\ (\triangle)$ and $\mu=30\ (\triangledown)$. The results
      for the same value of $\mu$ have been joined for clarity.
    }
    \label{fig:TCT}
\end{figure}

The dependence of the conductivity with various values of $\mu$ are
shown in \figno{TCT}.
Let us try to understand the physics behind the main characteristics
of the behavior of $\kappa$. In general, for low and high
temperatures, $\kappa$ increases, as in the FPU $\beta$ model
($\mu=0$). At high temperatures, $\kappa$ for different $\mu$ tends to
converge since the quartic coupling dominates the theory and the
effect of the quadratic pinning is relatively small. At low
temperatures, the theory approaches the harmonic limit and $\kappa$
diverges.  Differences due to $\mu$ are significant in this regime.

By obtaining $\kappa$ for various values of $L$ with respect to a
given set of $\mu,T$, the size dependence $\alpha$ may be obtained for
the particular set of parameters, $\mu,T$.
\begin{figure}[htbp]
    \centering
    \includegraphics[width=8cm,clip=true]{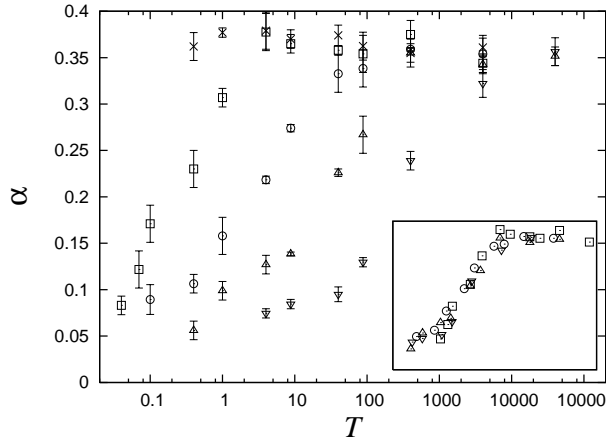}
    \caption{$\alpha$ vs. $T$ for $\mu=0\ (\times)$, $\mu=1\ (\Box)$,
      $\mu=4\ (\bigcirc)$, $\mu=10\ (\triangle)$ and $\mu=30\
      (\triangledown)$. 
      Inset shows the same points (for $\mu\not=0$) with the rescaled
      temperature $T/\mu^2$ (see text).}
    \label{fig:alphaT}
\end{figure}
The dependence of $\alpha$ on $T$ for various values of $\mu$ are
shown in \figno{alphaT}. The behavior of $\alpha$ can be summarized as
follows: At higher temperatures, $\alpha$ converges to one value and
it is consistent with the previous literature for the FPU $\beta$
model \cite{fpuRev,Lepri97,ak-fpu,Narayan02,Lepri03,Pereverzev,MDN07}.
When $\mu\not=0$, at lower $T$, the dependence $\alpha$ decreases with
$T$.  The crossover behavior to decreasing $\alpha$ occurs at larger
$T$ for larger $\mu$. This is natural since the effect of the pinning
term is greater for larger $\mu$ so that the crossover occurs at a higher
temperature.

\begin{figure}[htbp]
    \centering
    \includegraphics[width=8cm,clip=true]{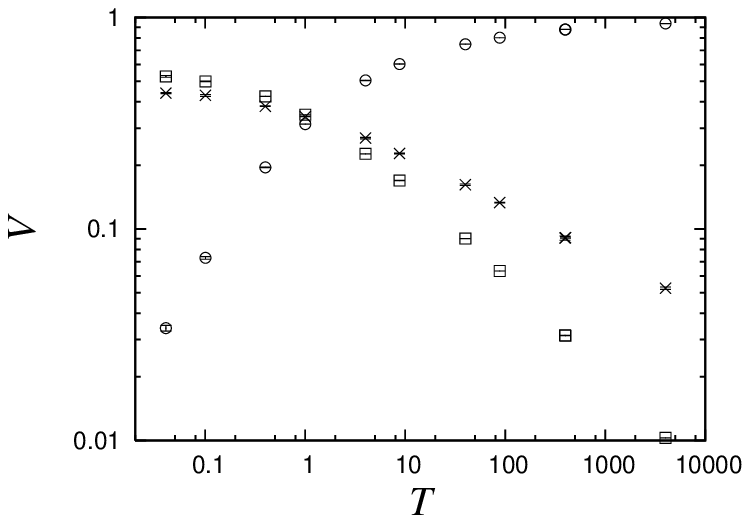}
    \caption{Temperature dependence of $\vev{\mu q^2}/T\ (\times)$,
      $\vev{(\nabla q)^2}/T\ (\Box)$ and $\vev{(\nabla q)^4}/T \
      (\bigcirc)$ for $\mu=1$.}
    \label{fig:crossover}
\end{figure}

The decrease in $\alpha$ can be understood as the effect of pinning,
with the tendency towards bulk behavior. 
This change in $\alpha$ occurs at low $T$ since this is the region
where the quadratic term is dominant and the quartic FPU coupling is
weak. 
The validity of the argument can be confirmed from the dominant term
in the potential, which indeed changes as we vary the temperature.
This crossover behavior can be investigated in equilibrium and an
example of this is seen in \figno{crossover}. The time averages of the
terms in the potential, ${\mu q_k^2}$, ${(\nabla q)_k^2}$ and
${(\nabla q)_k^4}$ will equal $T$, if the term by itself completely
dominates the potential, from the virial argument. We note here that
in the equilibrium system, the expectation values of ${\mu
  q_k^2}$, ${(\nabla q)_k^2}$ and ${(\nabla q)_k^4}$ do not
vary inside the system.
The above understanding of the crossover allows us to compute roughly
the crossover temperature, which we denote as $T_c$.
\begin{equation}
    \label{eq:naiveScaling}
    q^4\sim T\ (T\gg1),\qquad
    \mu q^2\sim  T\ (T\ll1) \quad\Rightarrow\quad
    T_c\sim \mu^2
\end{equation}
The estimate is consistent with the crossover temperatures seen in
\figno{alphaT}.  The picture also suggests that a more general scaling
with the rescaled temperature $T/\mu^2$ might apply. The behavior of
$\alpha$ with respect to this rescaled temperature is shown in the
inset and the scaling works well in practice. We do not have a
rigorous argument for this scaling behavior and it is worth further
study.  While there is visibly different behavior below and above this
``crossover temperature'', we expect this not to be a sharp
transition.
For $T\lesssim T_c$ and $\alpha$ not too small, the behavior seems to
be reasonably well described by
\begin{equation}
    \label{eq:alphaBehavior}
    \alpha=\alpha_0+\nu \ln {T\over T_c},\qquad \nu=0.055(5)
\end{equation}
Here, we denoted the common value of $\alpha$ high temperature
constant as $\alpha_0$.  If we naively extrapolate this behavior to
low temperatures, $\alpha=0$ is reached at a finite temperature,
$T_c\exp(-\alpha_0/\nu)$, but we cannot reliably extrapolate to this
limit.  Paradoxically, numerical determination of $\alpha$ is
exceedingly difficult for small $\alpha$ values, where we naively
expect the system to have bulk behavior.  This is because in the model
described by \eqnn{ham}, small $\alpha$ values occur close to the
harmonic limit where the gradients become small and are difficult to
compute reliably.
This difficulty precludes us from definitely predicting what happens
in the limit $\alpha\rightarrow0$.  Several scenarios are possible:
The model might reach $\alpha=0$ at finite $T$. In this case, if
$\kappa$ is finite, the system has bulk behavior.  It is also possible
that the behavior for small $T$ is not described by the log behavior
\eqnn{alphaBehavior} and that $\kappa$ diverges in the
$\alpha\rightarrow0$ limit, which is also the $T\rightarrow0$ limit
and this limit coincides with the harmonic model. In this case, the
model can, in some sense, be arbitrarily close to having bulk behavior
yet does not achieve it for finite values of the physical parameters.
We consider  this last possibility to be the most likely.
It can be noticed that the crossover temperature for the FPU $\beta$
model ($\mu=0$) is $T_c=0$ so that it is never reached, if the above
scaling is applied. However, the model is quite different from
$\mu\not=0$ cases so that it should be studied separately.

The behavior of the system as it becomes closer to the harmonic limit
perhaps needs to be explained. In the harmonic limit, one might expect
$\kappa\propto L$ behavior with $\alpha$ being one, so that $\alpha$
should rather increase than decrease when $T$ is smaller. This
behavior is suggested by the exact analytic solutions for the harmonic
models\cite{Rieder67,Nakazawa}; in these solutions, for fixed boundary
thermostat temperatures, $T_{1,2}^0$, $J$ is independent of $L$.
Naively, the gradient varies as $(T_2^0-T_1^0)/L$ so that $\kappa$
should be proportional to $L$.  However, this argument completely
ignores the boundary effects. With boundary jumps, which are
dynamical, the gradient is not directly related to the boundary
thermostat temperatures and $\alpha_J=1\gneq\alpha$. In fact, in the
harmonic case, the gradient is zero inside the system so that $\kappa$
effectively diverges even for a finite size system, which is
consistent with the behavior of $\kappa$ for small $T$ in our results,
as in \figno{TCT}.  Given the divergence of $\kappa$, there is no
analytic prediction for $\alpha$ in the harmonic limit. One property
of the harmonic model is the independence of $J$ with respect to $L$
for fixed boundary thermostat temperatures. We find that the
dependence of $J$ on $L$ does indeed become weaker at lower
temperatures. The dynamics of the interior adjusts the gradient so
that the conductivity has a weaker dependence on $L$.

In this work, we computed thermal transport properties of the FPU
$\beta$ model with a quadratic pinning potential for various values of
the pinning potential, temperature and system size. We have analyzed
the dependence of the conductivity on the physical parameters of the
theory and have obtained an understanding of them. Had we included the
quartic potential, the theory should have bulk behavior in some
parameter region\cite{WangLi06,Dhar07,LiLi07}. However, in the model
discussed, none of the couplings by themselves lead to bulk behavior
of the system and the total system does not reach the bulk limit.
At low temperatures, where the quartic coupling is effectively weak,
the quadratic pinning term becomes more important. This reduces
$\alpha$ (in $\kappa\propto L^\alpha$) so that the transport behavior
is closer to bulk behavior. However, the system governed by quadratic
terms is harmonic and the conductivity, at the same time, approaches
divergent behavior.

The size dependence, $\alpha$, varies with parameters of the theory.
We have used this power $\alpha$ in this study but it is worth noting
that a priori, there is no rigorous proof that $\kappa$ should behave
homogeneously as a power of $L$, even though we find that the behavior
applies quite well in practice.
Some questions remain: In particular, the precise limiting behavior of
the theory can not be obtained numerically, so that a rigorous
theoretical reasoning is quite desirable, particularly close to the
harmonic limit. To this end, one might envisage an application of
perturbation theory around the harmonic theory to the FPU quartic
coupling, similar in spirit to the perturbation theory applied to the
quartic pinning coupling \cite{Rafael}. From this perspective, one can
view the theory with the quadratic pinning in the FPU theory as the
extrapolation between the harmonic model with a continuous variation
of the size dependence of the thermal conductivity.

\end{document}